\documentclass{ws-ijmpa}
\usepackage[super,compress]{cite}
\usepackage[dvips,breaklinks]{hyperref}
\hypersetup{colorlinks,urlcolor=black,citecolor=black,linkcolor=black,filecolor=black}
\usepackage{breakurl}
\usepackage{graphicx}

\newcommand{\Eslash}{\mbox{$E \kern-0.6em\slash$}}
\newcommand{\pslash}{\mbox{$p \kern-0.5em\slash$}}
\newcommand{\met}{\mbox{\ensuremath{\Eslash_{\kern-0.3emT}\!}}}
\newcommand{\metvec}{\mbox{\ensuremath{\vec{\Eslash_{\kern-0.3emT}\!}}}}
\newcommand{\misspt}{\mbox{\ensuremath{\pslash_{\kern-0.1emT}\!}}}
\newcommand{\missptvec}{\mbox{\ensuremath{\vec{\pslash_{\kern-0.1emT}\!}}}}
\newcommand{\MET}{\ensuremath{\not\hspace*{-0.84ex}E_T\,}}

\begin{document}

%
\catchline{}{}{}{}{}
%

\title{REVIEW OF PHYSICS RESULTS FROM THE TEVATRON
}

\author{
D. Bandurin$^a$,
G. Bernardi\footnote{Editor for this
Special Issue of the International Journal of Modern Physics A}$^{^.,b}$ ,
C. Gerber$^c$,
T. Junk$^d$,
A. Juste$^e$,\\
A. Kotwal$^f$,
J. Lewis$^d$,
C. Mesropian$^g$,
H. Schellman$^h$,
J. Sekaric$^i$,\\
D. Toback$^j$,
R. Van Kooten$^k$,
C. Vellidis$^d$,
L. Zivkovic$^l$\\[0.5cm]
}

\address{{ 
$^a$
 University of Virginia, Department of Physics,
Charlottesville, Virginia 22904, USA\\
$^b$
LPNHE Paris, University of Paris VI \& VII, Paris 75005, France\\
$^c$
University of Illinois at Chicago 
Chicago IL 60607-7059, USA\\
$^d$
Fermi National Accelerator Laboratory
Batavia, Illinois 60510, United States of America\\
$^e$
Instituci´o Catalana de Recerca i Estudis Avan¸cats (ICREA) and
Institut de F´ýsica d'Altes Energies (IFAE)
Universitat Aut`onoma de Barcelona
E-08193 Bellaterra (Barcelona), Spain\\
$^f$
Department of Physics, Duke University,
Durham, NC 27708, USA\\
$^g$
The Rockefeller University,
1230 York Avenue
New York, NY 10065, USA \\
$^h$
Department of Physics, Northwestern University,
Evanston, IL 60208, USA\\
$^i$
Department of Physics and Astronomy, University of Kansas,
Lawrence, KS 66045, USA\\
$^j$
Mitchell Institute for Fundamental Physics and Astronomy,
Department of Physics and Astronomy,
Texas A\&M Univ.,
College Station, TX 77843, USA\\
$^k$
Department of Physics, Indiana University,
Bloomington, Indiana 47405, USA\\
$^l$
Laboratory for HEP, Institute of Physics Belgrade
Pregrevica 118, 11080 Zemun, Serbia
}}
\maketitle


\begin{abstract}
We present a comprehensive review of the physics results obtained by the
CDF and D0 collaborations up to summer 2014, with emphasis on those
achieved in the Run II of the Tevatron collider
which delivered a total integrated luminosity of $\sim$ 10 fb$^{-1}$ at $\sqrt{s}=1.96$ TeV.
The results are presented in six main physics topics: QCD, Heavy Flavor, Electroweak,
Top quark, Higgs boson and searches for New Particles and Interactions. The characteristics of the accelerator,
detectors, and the techniques used to achieve these results are also briefly summarized.
\end{abstract}

\ccode{PACS numbers:}


\tableofcontents

\section{Introduction}

Proton-antiproton ($p{\bar{p}}$)  collisions are ideal to study elementary particle collisions
at the highest energies, as was shown in the 1980's by the CERN $Sp\bar{p}S$ collider
at $\sqrt{s}$=630 GeV, provided enough luminosity (driven mainly by anti-proton intensity) can
be delivered.
The Fermilab Tevatron~\cite{tevbook}, a $p{\bar{p}}$ collider
with superconducting magnets, went a step further in energy and luminosity, and
operated from 1988 to 1996 at a center of mass energy of 1.8~TeV (Run I),
and from 2001 to 2011 at a center of mass energy of 1.96~TeV (Run II).
The CDF and D0 detectors each recorded approximately 0.1~fb$^{-1}$ of collision data
in Run~I, and approximately 10~fb$^{-1}$ of collision data in Run~II.
In the three years after the Tevatron's final shutdown in September 2011, the majority of the
analyses have been completed, although a few important results, like the
$W$ mass measurement and Top properties, with the full statistics, are still in preparation. 
It is thus appropriate to provide
now an almost comprehensive review of the main results, based on more than 900 published
papers, to reflect the legacy of the
Tevatron. A previous, shorter review was published in 2013~\cite{Grannis:2013xna}.
The emphasis in this review is on Run II results, but the comparison with a summary of Run I
results shows the progress which was achieved with a gain of two orders of magnitude
in luminosity at approximately the same center of mass energy.

The review is organized in seven chapters: one introductory chapter briefly summarizing
the experimental apparatus and the experimental techniques used, followed by six physics
chapters, devoted to QCD, Heavy Flavor, Electroweak,
Top quark, Higgs boson and searches for New Particles and Interactions.

\subsection{The Tevatron}

The Tevatron had its  first collisions in 1985, and Run 0 took place from 1988 to 1989.
The physics really started with Run I, which took place
from 1992 to 1996 and
 provided sufficient
collision data to produce a wealth of physics results. The center-of-mass energy of $\sqrt{s}$ = 1.8 TeV and a luminosity of up to 1.6
10$^{31}$ cm$^{-2}$ s$^{-1}$, provided an integrated luminosity of 0.1 fb$^{-1}$, and allowed
for the discovery of the top quark~\cite{Abe:1995hr, Abachi:1995iq}.

Extensive upgrades were
performed from 1996 to 2001 on the accelerator, the main ones being to replace the Main Ring with the Main Injector
in a new tunnel, and introducing a new
Recycler Ring for antiproton storage with electron cooling to further reduce the beam phase space.
Following these improvements, Run II began in 2001
at a center of mass energy of 1.96~TeV, and ended in 2011, when the Tevatron stopped operation
on September 30. In all physics topics, major
advances were achieved, as will be reviewed in the following chapters.
The Tevatron delivered to the CDF and D0
detectors
approximately 12~fb$^{-1}$ of collision data in Run~II, while
approximately 10~fb$^{-1}$ of excellent quality data were used in the ``full statistics'' analyses.

During Run~II, the Tevatron achieved a maximum instantaneous
luminosity of 4.31$\times$10$^{32}$~cm$^{-2}$s$^{-1}$
but the effect of the overlay of multiple interactions remained manageable.
Thirty-six bunches of protons collided with an equal number
of bunches of antiprotons with a bunch spacing of 396~ns, spaced with
gaps in order to allow kicker magnets to abort the beam cleanly.  The
luminous region had an RMS of approximately 25~cm along the beam
axis and approximately 30 $\mu$m in the directions transverse to the
beam axis
The Tevatron accelerator complex is discussed in
References \citen{Shiltsev:2013wna,Holmes:2011nq}.

\subsection{The CDF and D0 Detectors}

The main components of the CDF~\cite{Acosta:2006am,Abulencia:2007ix}
and D0~\cite{Abachi:1993em, Abazov:2005pn, Abolins:2007yz, Angstadt:2009ie} 
detectors
are the tracking, calorimeters and  muon detectors.
Here we briefly summarize their main characteristics,
together with their triggering
systems specific to Run II.

The kinematic properties of
particles and jets are defined with respect to the origin of the detector
coordinate system which is at the center of the detector.
To quantify polar angles the pseudorapidity  variable,
defined as $\eta = - \ln \tan (\theta/2$), is used  where $\theta$ is the polar angle
in the corresponding spherical polar coordinate system. Throughout this review
we use natural units, in which $c=\frac{h}{2 \pi}=1$.

\subsubsection{Tracking detectors}

The CDF tracking system consists of an eight-layer silicon microstrip tracker
and an open-cell drift chamber referred to as the central outer tracker (COT),
both immersed in a 1.4~T solenoidal magnetic field.  These combined systems
provide charged particle tracking and precision vertex reconstruction in
the pseudorapidity region $| \eta | <1.0$ with partial coverage in the COT
to $| \eta | < 1.7$ while the two outer layers
of the silicon detector extend the
tracking capability to $| \eta | < 2.0$.

The D0 tracking system 
consists of an inner silicon
microstrip tracker (SMT) surrounded by an outer central scintillating
fiber tracker (CFT). Both the SMT and CFT are situated within a 1.9 T
magnetic field provided by a solenoidal magnet surrounding the entire
tracking system.
The SMT is used for vertex
reconstruction and for tracking up to $|\eta| < 2.5$, generally in combination
with the CFT.
The CFT is also used for tracking and vertex reconstruction,
and provides precise tracking coverage up to $|\eta| < 1.7$.

\subsubsection{Calorimeters}

The CDF calorimeter system is used to measure the energy of charged and
neutral particles and 
can identify and measure photons, jets from partons, missing
transverse energy, and in combination with information from other systems electron
and tau leptons. They are arranged around the
outer edges of the central tracking volume and solenoid and consist of modular
sampling scintillator calorimeters with a tower-based projective geometry.
The inner electromagnetic sections of each tower consist of lead sheets
interspersed with scintillator, and the outer hadronic sections are composed
of scintillator sandwiched between sheets of steel. The CDF calorimeter
consists of two sections: a central barrel calorimeter and forward end plug
calorimeters covering the pseudorapidity region $ |\eta| <3.6$.

The D0 liquid-argon calorimeter system is used for the identification
and energy measurement of electrons,
photons, and jets, and also allows for the measurement of the missing transverse energy (\MET )
of the events, typically from undetected neutrinos.
It is located outside of the tracking and solenoid systems.
The central calorimeter (CC) covers detector pseudorapidities $|\eta| \le 1.1$
and two additional end-cap calorimeters extend the range up to $|\eta| = 4.2$.
These fine-grained calorimeters are subdivided into electromagnetic (EM) followed by fine hadronic and
then coarse hadronic sections.
The intercryostat plastic scintillator detectors 
complete the calorimeter coverage in the intermediate pseudorapidity region $0.8<|\eta|<1.4$.

\subsubsection{Muon detectors}

The CDF muon detector is made up of four independent detector systems outside
the calorimeter modules and consists of drift chambers interspersed with
steel layers to absorb hadrons. The central muon detector (CMU) is mounted
directly around the outer edge of the central calorimeter module and detects
muons in the  pseudorapidity region $| \eta | <0.6$.  The central muon
extension is composed of spherical sections and extends the pseudorapidity
coverage  in the range $0.6< | \eta | <1.0$.  The central muon upgrade (CMP) surrounds portions of the CMU
and central muon extension (CMX) systems covering gaps in angular coverage and allowing excellent
identification of higher
momentum muons due to additional layers of steel absorber.
The barrel muon upgrade (BMU)
is a barrel shaped extension of the muon system in the pseudorapidity
region $1.0< | \eta | <1.5$.  The CMX,
CMP and BMU systems also include
matching scintillator systems which provide timing information to help
identify collision produced muons.

The D0 muon detector system consists of a central muon detector system
covering the range $|\eta|<1$ and  forward muon systems which cover
the region $1<|\eta|<2$.
Both central and
forward systems consist of a layer of drift tubes and scintillators
inside toroidal magnets and two similar
layers outside the toroids.
Scintillation counters are included for triggering purposes
and the 1.8 T toroidal iron magnets make it possible to determine muon momenta and
perform tracking measurements within the muon system alone, although in general the
central tracking information is also used for muon reconstruction.

The D0 tracking system benefits from a unique feature among high energy collider
detectors, namely the ability to reverse regularly (typically every two weeks)
the polarities of the solenoid and of the toroid magnets, providing data in 
four polarity combinations. This allows for a significant reduction of experimental
systematic uncertainties related to charged particle properties and for measurements 
achieving excellent precision, for instance the measurement of inclusive like-sign
dimuon asymmetry described in the heavy-flavor physics chapter.

\subsubsection{Triggering systems}

The CDF trigger system consists of three levels.  The level one trigger
consists
of dedicated electronics that operate at the beam crossing frequency.
This system can identify and measure the transverse momentum of
charged particles using COT information
which provides the basis of several
trigger decision criteria.  This information is also combined
with information from the calorimeters or
muon systems to provide a trigger for leptons.  The calorimeter
trigger hardware measures energy clusters which are used to identify  jets and photons
as well as an imbalance in event transverse energy interpreted as \MET.  
The second-level trigger operates with a mixture of hardware and
software algorithms.  It refines the measurements made by
the level one trigger and uses broader combinations of information
from different subsystems.  It adds precision tracking information
from the silicon detectors to the fast COT tracks to form trigger
decisions sensitive to the presnce of displaced vertices.

The D0 trigger system also has three
trigger levels referred to as L1, L2 and L3. Each consecutive
level receives a lower rate of events for further
examination.
The L1 hardware based elements of the
triggers used in the electron channel typically require calorimeter
energy signatures consistent with an electron. This is  expanded at L2
and L3 to
include trigger algorithms for instance 
requiring an electromagnetic object together with at least
one jet for which the L1 requirement is calorimeter energy depositions consistent with
high-$p_{T}$ jets.
For some inclusive muon samples, events are triggered using
the logical {\tt OR} of the full list of available triggers
of the D0 experiment. The muon trigger pseudorapidity coverage is restricted
to  $|\eta| < 1.6$ where the majority of  $W\mu\nu+$jet events ($\simeq$65\%) are
collected by triggers requiring high-$p_{T}$ muons at L1.
Events not selected by the high-$p_{T}$ muon triggers
are primarily collected by jet triggers.

\subsection{Physics object identification at the Tevatron}

\subsubsection{Lepton and photon identification}

Isolated electrons and photons are reconstructed in the calorimeter
and are selected
in the pseudorapidity regions
$|\eta|<1.1$ at CDF, and at
$|\eta|<1.1$ and $1.5 < |\eta| < 2.5$ at D0~\cite{Abazov:2013tha}.
The EM showers are required to pass spatial distribution requirements consistent
with those expected from electrons for each section of the calorimeter, and D0 benefits
in this respect from the fine longitudinal segmentation of its calorimeter.
For electron identification, in CDF a matching track is required within the coverage of the COT tracker,
while in the D0 CC region, a reconstructed track, isolated from other tracks, is also required to be matched
to the EM shower.

Muons are selected by requiring
a local track spanning all layers of the muon detector system (for D0 both within as well as outside of the
toroidal magnet).  A spatial match is then required to a corresponding track
in the COT (CDF) or CFT (D0)~\cite{Abazov:2013xpp}.
To suppress muon background events originating from the semileptonic
decay of hadrons, muon candidate tracks are required to be separated from
jets by at least $\Delta R = \sqrt{(\Delta \eta)^{2} + (\Delta \varphi)^{2}} > 0.5 $.
A veto against cosmic ray muons is also applied using scintillator timing information in D0 and
a specialized tracking algorithm is used both at CDF and D0 to track cosmic ray muons
passing through both sides of the detector.
Muons can also be identified as a minimum ionizing isolated track in CDF for regions without
muon coverage taking advantage of the fact that muons interact with
low probability in the material of the calorimeter leaving only a
small ionization signature.

Multivariate algorithms
are used to enhance efficiency and background rejection
in some electrons- and muon-based analysis.

Tau lepton decays into hadrons are characterized as narrow, isolated jets with
lower track multiplicities than jets originating from quarks and gluons. Three types of tau lepton
decays are distinguished by their detector signature. The type-1 category comprises
one-track tau decays
consisting of energy deposited in the hadronic calorimeter associated with a
single track; type-2 corresponds to one-track tau
decays with energy deposited in both the hadronic and EM calorimeters; 
type-3 are multitrack decays with
energy in the calorimeter and two or more associated tracks with invariant mass
below 1.7 GeV.
In D0, a set of artificial neural networks (NN), one for each tau type, is applied
to discriminate hadronic tau decays from jets. The input variables are
related to isolation and shower shapes, and exploit correlations between
calorimeter energy deposits and tracks. When requiring the neural network
discriminants to be above thresholds optimized for each tau type separately,
typically  65\% of taus are retained, while 98\% of the multijet (MJ) background is
rejected in the kinematic phase space of the analyses.
In CDF boosted decision tree based algorithms are used for the same purpose.

\subsubsection{Jets, $b$ and $c$ jets, and missing transverse energy}

Jets are reconstructed in the calorimeters both at CDF and D0.
In CDF, for QCD physics, jets are generally reconstructed with a mid-point cone jet algorithm 
with a cone of size $\Delta R < 0.7$. Otherwise, 
jets are reconstructed using a calorimeter based clustering algorithm,
with a cone of size $\Delta R < 0.4$.
In D0, jets are reconstructed 
for  $|\eta| < 3.2$ using the D0 Run II mid-point cone algorithm~\cite{Abazov:2013hda}.
Calorimeter energy deposits
within a cone of size $\Delta R < 0.7 $ are used to form the jets in QCD
physics measurements, otherwise a cone of size 0.5 is used.
The energy of the jets is calibrated by applying
a jet energy scale correction determined using $\gamma$+jet and $Z$+jet events~\cite{Bhatti:2005ai,Abazov:2013hda}.

Jet identification  efficiency and jet resolutions are adjusted in the simulation to match those measured in data.

At both CDF and D0, the identification of quarks initiated by
a $b$ or $c$ quark (``$b$ or $c$ tagging'') is done in two steps~\cite{Abazov:2010ab,Abazov:2013gaa}.
The jets are first required to pass the taggability requirement based on charged particle
tracking and vertexing information, to ensure that they originate from the interaction vertex and
that they contain charged tracks.
At CDF the next step in $b$ tagging is done using a NN with similar variables but including additional track
quality information~\cite{Freeman:2012uf}.  The CDF experiment also employs a cut based secondary
vertex tagger.
At D0  a $b$ tagging NN is
applied to the taggable jets. This NN uses a combination
of seven input variables, five of which contain secondary vertex information; the
number and mass of vertices, the number of and $\chi^{2}$ of the vertex contributing
tracks, and the decay length significance in the $x-y$ plane. Two impact parameter
based variables are also used.
As an example at D0 the typical efficiency for identifying a $p_T = 50~\rm GeV$ jet
that contains a $b$ hadron is $(59\pm 1)$\% at a corresponding misidentification rate of 1.5\%\ for light
parton ($u,d,s,g$) initiated jets. This operating point is typically used for events with two
``loose''  $b$-tagged jets.
When tightening the identification requirement, the efficiency for identifying a jet
with $p_T$ of 50~GeV
that contains a $b$ hadron is $(48\pm 1)$\% with a misidentification rate of 0.5\%\ for
light parton jets. Additional requirements are put for $c$ jet identification, and these are
typically analysis dependent so are not described here. Typical
final identification efficiency for $c$ jets are around 20\%, for a misidentification rate
of a few \%.

The event missing transverse energy (\MET )is calculated
from individual calorimeter cell energies in the calorimeter.
It is corrected for the presence in the event of any muons and
all energy corrections to EM objects or to the jets  are propagated to \MET .
Both experiments identify events with instrumental \MET\ by comparing missing
transverse energy calculations based on either reconstructed tracks or
calorimeter deposits.  The CDF
experiment also employs an algorithm that combines tracking and calorimeter
information to improve \MET resolution.

\vspace{2cm}

\noindent
{\bf On the arXiv, the following chapters, Quantum Chromodynamics Studies,
Heavy Flavor physics, Electroweak Physics,Top Quark Physics,Higgs Boson Physics,
and Searches for New Particles and New Interactions, are posted as separate
arXiv submissions}

\vspace{2cm}

\noindent
{\bf Acknowledgments:} acknowledgements are provided in the physics chapters.

\section{Quantum Chromodynamics Studies at the Tevatron}
	\subsection{Run I results}
	\subsection{Jet final states}
	\subsection{Photon final states}
	\subsection{W/Z+jets final states}
	\subsection{Soft QCD}

\section{Heavy Flavor physics at the Tevatron}
	\subsection{Characteristics of Flavor Physics at the Tevatron and Run I Results}
	\subsection{Production}
	\subsection{Spectroscopy}
	\subsection{Decays/Lifetimes}
	\subsection{Mixing and Oscillations of Heavy Neutral Mesons}
	\subsection{CP Violation (CPV)}
	\subsection{Rare Decays}

\section{Electroweak Physics at the Tevatron}
	\subsection{{Run I Results}}
	\subsection{Physics of the $W$ and $Z$  vector bosons}
	\subsection{Precision Electroweak Measurements on the $W$ boson}
	\subsection{Diboson Production at the Tevatron}

\section{Top Quark Physics at the Tevatron}
	\subsection{The Top Quark in Run I}
	\subsection{Studies of Top Quark Pair Production}	
	\subsection{Observation and Studies of Single Top Quark Production}
	\subsection{Top Quark Mass}
	\subsection{Top Quark Properties}

\section{Higgs Boson Physics at the Tevatron}
	\subsection{Higgs Boson Theory and Phenomenology}	
	\label{sec:sec2}
	\subsection{Analysis Tools}
	\label{sec:sec4}
	\subsection{Searches for the Standard Model Higgs Boson}
	\label{sec:sec5}
	\subsection{Searches for Higgs Bosons Beyond the Standard Model}
	\label{sec:sec6}

\section{Searches for New Particles and Interactions at the Tevatron}
	\subsection{{Theoretical Motivation}}\label{sec_theory}
	\subsection{{Run I Results}}
	\subsection{{Supersymmetry}}\label{sec_susy}
	\subsection{{Other BSM Searches}}\label{sec_nonsusy}

\section{Summary}

\section*{Acknowledgments}

\newpage

\bibliographystyle{ws-ijmpa}
\bibliography{Tevatron_Review}
\end{document}